\documentclass[preprint,12pt]{elsarticle}
\usepackage{amsmath}
\usepackage{amsfonts}
\usepackage{amssymb}
\usepackage[utf8]{inputenc}
\usepackage[T2A]{fontenc}
\usepackage{bm}

\begin{document}
	\begin{frontmatter}
		\title{
		General solution of the Dirac equation \\for electrons bound by a charged atomic chain
		}	
		
		\author{Alexander Eremko\fnref{fn1}} 
		\author{Larissa Brizhik\fnref{fn2}\corref{cor1}} 
		\author{Vadim Loktev\fnref{fn3}}
	
	\address{Bogolyubov Institute for Theoretical Physics of the National Academy of Sciences of Ukraine \\
		Metrologichna Str., 14-b,  Kyiv, 03143, Ukraine}
		
		
		\fntext[fn1]{eremko@bitp.kyiv.ua}
		\fntext[fn2]{brizhik@bitp.kyiv.ua}
		\fntext[fn3]{vloktev@bitp.kyiv.ua}
		\cortext[cor2]{Corresponding author}


\begin{abstract}
{The system of electrons bound by a charged atom chain is studied within the Dirac theory.  The general analytical solution of the Dirac equation is obtained. Analytical expression for electron  energy is derived from which it follows that the principal quantum number can be introduced for electron states in such a system, similar to the case of a hydrogen atom. We prove that electrons bound by the atomic chain, are fully collectivized and the energy  can be calculated as a function of the occupation number. The spectral band with the
	principal quantum number $n$ is shown to be split into $n$ subbands forming the fine structure. The scale
	of the fine structure splitting is calculated.}	
\end{abstract}

\maketitle

\end{frontmatter}

\section{Introduction} 

Rapid development of advanced methods of nanomaterial fabrication down to the atomic scale  has been achieved over the past few decades. Among such materials are single-metal-atom chains (SMACs), free standing SMACs, such as Au, Pt, Ir, Mn, and Fe monoatomic chains and other. The review \cite{MingyuMaCR2022} on structure, synthesis, and properties of SMACs shows how fast this research field has been growing during the last decade. They are  model systems for experimental study of the unique physical and chemical properties of one-dimensional (1D) systems  \cite{LinNanoLet2017,QinNatElectr2020,ComtetNat2019}. On the other hand, a monatomic chain is a fundamental theoretical model in Solid State Physics for description of the dispersion relations for phonon and electron energies in 1D lattices of identical atoms. Such models have been also widely used to investigate complex phenomena
in many-body systems such as the Tomonaga-Luttinger liquid (TLL) \cite{LuttingerJMatPh1963}. Their reduced dimension enables exact solutions of microscopic Hamiltonians for the TLL and Hubbard models. 

It is known that due to confinement of electron motion to a 1D along an atomic chain, SMACs show many unique phenomena, including quantized ones, e.g., quantum effects of thermal conductance through atomic chains \cite{OzpineciPhRe2001}. Carbon atomic chains show huge dimerisation ratio \cite{LinNanoLet2017}, doped 1D cuprate chains demonstrate anomalously strong near-neighbour attraction \cite{ZhuoyuSci2021} and extreme anharmonicity \cite{CignarellaNanoL2025}, strong quantum confinement \cite{YaZhaoCompMat2026}, ultrahigh electron mobility \cite{LiApplPhL2025} as well as many other unusual properties (see review paper \cite{MingyuMaCR2022}). In addition to this, a variety of non-trivial quantum effects such as the so-called spin-charge separation and Majorana fermions, have been theoretically proposed and experimentally observed in materials with 1D structural motif \cite{ZhuoyuSci2021,WangNat2022,SchneiderNatNano2022,NadjSci2014}. 

It is worth to emphasize that SMACs have attracted much attention not only due to their unique properties, but also due to their numerous promising applications in electronics, spintronics and beyond \cite{ZhangMatTodayPh2024,LiaoJSemicond2025}. This requires fundamental understanding of the properties of electrons squeezed into 1D structures taking into account the fact that electrons move in the common 3D space and acquire a 1D behavior due to their confinement by the Coulomb field of the chain atoms (ions). An atomic chain of $ N \gg 1 $ atoms with the charge $ Q = Ze $ ($ e $ is the elementary charge), regularly placed along $ z $ axis, creates the potential 
\begin{equation}
	\label{pot}
	V\left( \mathbf{r} \right) = - \sum_{n} \frac{eQ}{\sqrt{x^{2} + y^{2} + (z - na)^{2}}}  .
\end{equation}
which has the translational symmetry along the chain: $ V\left( \mathbf{r} + na\mathbf{e}_{z} \right) = V\left( \mathbf{r} \right) $. Here $ \mathbf{r}_{n} = \lbrace 0, 0, na \rbrace $ ($ n = 1,2,\ldots \,, N $) is the radius-vector of the $n$-th atom and $ a $ is the lattice spacing. This potential can be represented in another form 
\begin{equation}
	\label{V(r)}  
	V\left( \mathbf{r} \right) = \sum_{g_{n}} A_{g_{n}} \left( \mathbf{r}_{\perp} \right) e^{ig_{n}z}
\end{equation}
where $ g_{n} = 2\pi n/a $ with $ 2\pi/a $ is a reciprocal length of a 1D lattice, and the coefficients $ A_{g_{n}} \left( \mathbf{r}_{\perp} \right) $ are given by expressions 
\begin{equation}
	\label{a_g} 
	A_{g_{n}} \left( \mathbf{r}_{\perp} \right) = \frac{1}{a} \int_{(\text{over cell})} V\left( \mathbf{r} \right) e^{-ig_{n}z} dz = -\frac{eQ}{a} \int_{-a/2}^{a/2} \frac{e^{-ig_{n}z}}{\sqrt{\mathbf{r}_{\perp}^{2} + z^{2}}} dz.  
\end{equation} 
Thus, the  potential created by the periodic atomic chain, can be represented as $ V \left(\mathbf{r} \right) = \sum_{g_{n}} V_{g_{n}} \left( \mathbf{r}_{\perp},z \right) $. According to Eq. (\ref{a_g}),  Fourier coefficients $ A_{g_{n}} $ and, therefore, $ V_{g_{n}} $ decrease with reciprocal length  $ g_{n} $increasing. 

To develop a theory of electron states in periodic 1D crystals, it is useful to consider as the first step the quantum mechanical solution of the problem with the potential $ V ( \mathbf{r}) $ given by the zero-order Fourier component of the potential 
\begin{equation} 
	\label{V_0} 
	V_{0} \left( \mathbf{r}_{\perp} \right) = -\frac{eQ}{a} \int_{-a/2}^{a/2} \frac{dz}{\sqrt{\mathbf{r}_{\perp}^{2} + z^{2}}} 
	= \frac{2eQ}{a} \ln \left( \sqrt{ 1 + \dfrac{1}{4} \frac{a^{2}}{r_{\perp}^{2}}} - \frac{1}{2} \frac{a}{r_{\perp}} \right) ,
\end{equation} 
only. This expression, in which $ r_{\perp} = |\mathbf{r}_{\perp}| = \sqrt{x^{2} + y^{2}} $, describes the potential of a charged string with the linear charge density $ Q/a $. Electrons bound by such a string, possess 1D properties, and the corresponding solutions describe "free quasi-1D electrons". Similar to the 3D case, these solutions can	 be used for constructing the Bloch functions and band structure of the electron spectra in a periodic potential of a 1D crystal.

Usually theoretical studies of 1D systems have been done within the nonrelativistic scheme, based on the Shr\"{o}dinger equation or spin exchange Hamiltonians, like $ tJ $ models. Meantime it is known that in some 1D systems it is necessary to take into account relativistic corrections. In particular, it has been found  that the transition metals Ir, Pt, and Au form stable monatomic chains namely in the result of relativistic effects  \cite{CalvoPhRevLet2018}. Also  the intrinsic strong spin-orbit coupling (SOC) in the quasi-1D spin chains containing $ 4f $ rare-earth ions has  been reported \cite{SchafferRefPrPhys2016,LiuCninPhLet2018}. 

Note that SOC operator appears in the Shr\"odinger (and/or Pauli) Hamiltonian as the relativistic correction in the first approximation of the Dirac equation (DE) for the non-relativistic energy. This term, known as the relativistic Thomas correction, has different forms depending on the geometry of the considered problem -- well known spin-orbit interaction in the radial field in atoms,  Rashba and/oor Dresselhaus spin-orbit interaction in quasi-2D structures, etc. Usually the SOC operator as a small correction is taken into account using perturbation theory. It is not evident if such assumption is valid in the general case for any system. Compared to the 3D systems, the intrinsic SOC  increases with dimensionality decreasing. In particular, the so called giant Rashba  effect takes place in quasi-2D layered structures \cite{CrepaldiPhRevLet2012}, also strong SOC in 1D spin chains has been reported \cite{SchafferRefPrPhys2016}. 

It is worth to stress that the correct description of spin properties is provided by the DE which in some cases admits an exact solution. In Dirac theory coupling of the spin and spatial variables naturally appears in the exact solutions of the DE which does not include any special SOC term. This allows to consider the non-relativistic energy approximation more strictly. 

Taking these facts into account, in the present paper we study electrons bound by a charged atomic chain within the Dirac theory. So far to our knowledge this is the first time an atomic chain is investigated based on the DE in the general approach without specifying a particular spinor invariant. This is generalisation of the model \cite{BEL1-DI} where this problem ws solved with the choice of the relativistic spin polarization invariant.  We find the general analytical solution of the DE for such a system and show that it contains free parameters which play the role of spinor variables. We obtain the analytical expression for the energy, which shows that, as in the case of a hydrogen atom, the principal quantum number can be introduced to characterise the electron states. We prove that electrons bound by the atomic chain, are fully collectivized and the energy as a function of the occupation number can be calculated. 

\section{Dirac equation} 

We remind here again that the most adequate and complete description of electron states, naturally including the spin degree of freedom, is provided by the DE 
\begin{equation}
	\label{DE}
 \hat{H}_{D} \Psi = E \Psi 
\end{equation}  
where the Dirac Hamiltonian $ H_{D} = c \bm{\hat{\alpha}} \cdot \hat{\mathbf{p}} + m c^{2} \hat{\beta} + V \left(\mathbf{r} \right) $ is presented through the Dirac $ 4\times 4 $ matrices $ \hat{\beta} $ and $ \hat{\alpha}_{j} $ ($ \bm{\hat{\alpha}} = \sum_{j} \mathbf{e}_{j} \hat{\alpha}_{j} $ with $ j=x,y,z $). Usually its $ 2\times 2 $ block form is used:
\begin{equation}
	\label{H_D-m} 
	\hat{H}_{D} = \left( \begin{array}{cc}
		\left( V\left( \mathbf{r} \right) + mc^{2} \right) \hat{\mathbb{I}}_{2} & c\bm{\hat{\sigma}} \cdot \mathbf{\hat{p}} \\
		c\bm{\hat{\sigma}} \cdot \mathbf{\hat{p}} & \left( V\left( \mathbf{r} \right) - mc^{2} \right) \hat{\mathbb{I}}_{2}
	\end{array}  \right) .
\end{equation} 
Here $ \hat{\mathbb{I}}_{2} $ is a unit $ 2\times 2 $ matrix, $ \bm{\hat{\sigma}} = \sum_{j} \mathbf{e}_{j} \hat{\sigma}_{j} $ where $ \hat{\sigma}_{j} $ are Pauli matrices, and $ V \left(\mathbf{r} \right) $ is the potential created by the charged monatomic chain. As it was mentioned in the Introduction, here we consider the potential $ V \left(\mathbf{r} \right) = V_{0}(r_{\perp}) $ given by its zero-order Fourier component (\ref{V_0}) only. We approximate it by the main term of expansion with respect to ratio $ a/r_{\perp} $,
\begin{equation}
	\label{V_eff} 
	V_{eff} = - \frac{C_{eff}e^{2}Z}{r_{\perp}}  
\end{equation} 
which is the potential (\ref{V_0}) of the charged string in the limit $ a \rightarrow 0 $. 

Stationary states are characterized by the set of quantum numbers which reflect the eigenvalues of the complete set of commuting operators. This set includes the Hamiltonian operator and constants of motion (or invariants). The latter in the potential of the charged string $ V_{0}\left( \mathbf{r}_{\perp} \right) $ with the translational symmetry along $ z $-axis are $ z $-components of the momentum $ \hat{p}_{z} $ and of the total angular momentum $ \hat{J}_{z}$ which are connected with spatial degrees of freedom. Existence of the electron spin requires an additional spin invariant. Recall, for a charge motion in a free 3D space ($ V = 0 $) there are many scalar and vector spinor invariants, such as the components of the vector operators 
\begin{equation}
	\label{mu}
	\bm{\hat{\mu}} = \bm{\hat{\Sigma}} + \frac{\bm{\hat{\Gamma}} \times \mathbf{p}}{mc} ,
\end{equation} 
\begin{equation}
	\label{eps}
	\bm{\hat{\epsilon}} = \bm{\hat{\Omega}} \times \mathbf{p} , 
\end{equation} 
\begin{equation}
	\label{genS}
	\mathbf{\hat{S}} = \bm{\hat{\Omega}} + \hat{\rho}_{1} \frac{\mathbf{p}}{mc} , 
\end{equation}
and scalar operator 
\begin{equation}
	\label{S_0}
	\hat{S}_{0} = \bm{\hat{\Sigma}} \cdot \hat{\mathbf{p}} .
\end{equation} 
Here the vector operator (\ref{mu}) is  treated as the magnetic spin polarization and operator (\ref{eps}) as the electric spin polarization. The 3D vector (\ref{genS}) and the helicity operator (\ref{S_0}) are determined as the components of the 4-pseudovector of spin polarization \cite{Kessler}. 

In the potential of the charged string (\ref{V_eff}) when a free motion is allowed only along $ z $ axis and the constant of motion is the momentum $ \hat{\mathbf{p}} = \hat{p}_{z} \mathbf{e}_{z} $, not all spinor operators (\ref{mu})-(\ref{S_0}) are integrals of motion. In this case only three components of the magnetic spin polarization (\ref{mu}), two components of the electric spin polarization (\ref{eps}), one component of the spin polarization (\ref{genS}), and the helicity operator (\ref{S_0}) are invariants. 

Spin polarization operator $ \hat{\mathcal{S}}_{z} $ was used as a spinor invariant in \cite{BEL1-DI} for description of "free quasi-1D electrons" confined by a charged string. But this choice is only one among  others described above. Moreover, an additional integral of motion was found in  \cite{BEL1-DI} for the DE with the field $ Q/r_{\perp} $, which increases the number of non-commuting spinor invariants. 

Each spinor invariant takes two eigenvalues of the opposite sign. This means that together with the other quantum numbers, the eigenstates are defined also by the sign of the corresponding spinor invariant eigenvalue, $ \sigma = \pm 1 $. Each  complete set of commuting operators results in its own system of the eigen bispinors with the set of quantum numbers which include $ \sigma $. Vector spin operator $ \mathbf{s} = (\hbar/2) \bm{\hat{\Sigma}} $ is not a constant of motion (even in the non-relativistic limit) and spin orientation can be calculated as the quantum-mechanical meanvalue in the given eigenstate $ \psi_{\sigma} $ 
\[ 
\langle \mathbf{s} \rangle_{\sigma} = 
\langle \psi_{\sigma} \vert (\hbar/2) \bm{\hat{\Sigma}} \vert \psi_{\sigma} \rangle = (\hbar/2) \sum_{j = x,y,z} \mathbf{e}_{j} \langle \psi_{\sigma} \vert \hat{\Sigma}_{j} \vert \psi_{\sigma} \rangle . 
\]
Direct calculations give $ \langle \mathbf{s} \rangle_{-\sigma} = -\langle \mathbf{s} \rangle_{\sigma} $ which shows that in the  states $ \sigma = +1 $ and $ \sigma = -1 $ the spin is orientated in opposite direction. Therefore, two-valued number $ \sigma = \pm 1 $ can be treated as the spin quantum number. Spatial orientation of this direction (so called spin quantization axis) depends on the particular choice of the spinor invariant.

All this testifies that  "free quasi-1D electrons" have some spin freedom similar to 3D free electrons. It means that for modelling free 1D electrons we can use the general solution of the DE with the potential (\ref{V_0}) without fixing the spinor invariants. This solution contains some arbitrary parameters which play the role of the spin variables. 

According to the $ 2\times 2 $ block form presentation of the Dirac matrices, below we will use the Dirac bispinors $ \Psi \left( \mathbf{r} \right) $ in the form 
\begin{equation}
	\label{bispinor} 
	\Psi \left( \mathbf{r} \right) = \left( \begin{array}{c} \psi^{(u)} \left( \mathbf{r} \right)  \\ \psi^{(d)} \left( \mathbf{r} \right) 
	\end{array} \right) , \quad 
	\psi^{(u)} = \left( \begin{array}{c} 
		\psi^{(1)} \left( \mathbf{r} \right)  \\ \psi^{(2)} \left( \mathbf{r} \right) 
	\end{array} \right) , \; \psi^{(d)} = \left( \begin{array}{c}
		\psi^{(3)} \left( \mathbf{r} \right)  \\ \psi^{(4)} \left( \mathbf{r} \right) 
	\end{array} \right) , 
\end{equation}
where $ \psi^{(u/d)} $ are its upper/lower spinors, respectively, with the components $ \psi^{(\nu )} $ and $ \psi^{(\nu + 1)} $  ($ \nu =1 $ for the upper bispinor, and $ \nu = 3 $ for the lower one). 

Therefore, the DE (\ref{DE}) is the system of two equations for the upper and lower spinor components of the bispinor (\ref{bispinor})
\begin{equation}
	\label{DsysEq_1} 
	\begin{array}{c} 
		c\bm{\hat{\sigma}} \cdot \mathbf{\hat{p}} \psi^{(u)} + 
		\left( V_{0}(r_{\perp}) - mc^{2} - E \right) \psi^{(d)} = 0 , \\ 
		c\bm{\hat{\sigma}} \cdot \mathbf{\hat{p}} \psi^{(d)} +  
		\left( V_{0} (r_{\perp}) + mc^{2} - E \right) \psi^{(u)} = 0  
	\end{array}
\end{equation} 
with the potential $ V_{0}(r_{\perp}) $ defined in Eq. (\ref{V_eff}). 

Because $ z $-components of the momentum $ \hat{p}_{z} $ and total angular momentum $ \hat{J}_{z} = \hat{L}_{z} \hat{\mathbb{I}} + (\hbar/2) \hat{\Sigma}_{z} $ are constants of motion, the eigen bispinor of the DE satisfies equations $ \hat{p}_{z} \Psi = \hbar k \Psi $ and $ \hat{J}_{z} \Psi = \hbar M \Psi $ where $k\equiv k_{z} $ is the wave number along the chain, and the half-integer number $ M = \pm 1/2, \pm 3/2 , \ldots $ is the eigenvalue of the operator $J_z$. Due to the symmetry of $ V_{0} $, it is naturally to use cylindrical coordinate system in which the radius-vector and momentum operator can be represented as $ \mathbf{r} = \mathbf{r}_{\perp} + z \mathbf{e}_{z} $ and $ \hat{\mathbf{p}} = \hat{\mathbf{p}}_{\perp} + \hat{p}_{z} \mathbf{e}_{z} $ with $ \mathbf{r}_{\perp} = x \mathbf{e}_{x} + y \mathbf{e}_{y} $ and $ \hat{\mathbf{p}}_{\perp} = \hat{p}_{x} \mathbf{e}_{x} + \hat{p}_{y} \mathbf{e}_{y} $. Hence, the kinetic energy operator in Eqs. (\ref{DsysEq_1}) is $ \bm{\hat{\sigma}} \mathbf{\hat{p}} = \bm{\hat{\sigma}} \mathbf{\hat{p}}_{\perp} + \hat{\sigma}_{z} \hat{p}_{z} $, and orbital momentum  $ z $-component is $ \hat{L}_{z} \mathbf{e}_{z} = \mathbf{r}_{\perp} \times \hat{\mathbf{p}}_{\perp} = \left( x \hat{p}_{y} - y \hat{p}_{x} \right) \mathbf{e}_{z} $ and $ \hat{L}_{z} = -i\hbar \partial /\partial \varphi $.

It follows from eigenvalue equations for the two invariants $ \hat{p}_{z} $ and $ \hat{J}_{z} $ that the upper and lower spinor components of the eigen bispinor with the quantum numbers $ k,M $ are 
\begin{equation}
	\label{comps1} 
	\psi_{k,M}^{(u/d)} = e^{i \left( kz + M\varphi \right) } \chi^{(u/d)}  , \quad 
	\chi^{(u)} = \left( \begin{array}{c} 
		e^{-i\varphi /2} f_{1} (\mathbf{r}_{\perp}) \\ 
		e^{i\varphi /2} f_{2} (\mathbf{r}_{\perp}) \end{array} \right) , \quad 
	\chi^{(d)} = \left( \begin{array}{c} 
		e^{-i\varphi /2} f_{3} (\mathbf{r}_{\perp}) \\ 
		e^{i\varphi /2} f_{4} (\mathbf{r}_{\perp}) \end{array} \right) 
\end{equation} 

To write down equations in the cylindrical coordinate system, it is convenient to introduce the Pauli matrices corresponding to the basic vectors $ \mathbf{e}_{\mathbf{r}_{\perp}} $, $ \mathbf{e}_{\varphi} $, $ \mathbf{e}_{z} $ in this system: 
\begin{equation} 
	\label{sigma_rho_phi_z} 
	\hat{\sigma}_{\mathbf{r}_{\perp}} = \bm{\hat{\sigma}} \cdot \mathbf{e}_{\mathbf{r}_{\perp}} = \left( 
	\begin{array}{cc}
		0 & e^{-i\varphi} \\ e^{i\varphi} & 0
	\end{array} \right) , \:
	\hat{\sigma}_{\varphi} = \bm{\hat{\sigma}} \cdot \mathbf{e}_{\varphi} = 
	\left( \begin{array}{cc}
		0 & -i e^{-i\varphi} \\ i e^{i\varphi} & 0
	\end{array} \right) , \:
	\hat{\sigma}_{z} = 
	\left( \begin{array}{cc}
		1 & 0 \\ 0 & -1
	\end{array} \right) .
\end{equation} 
They satisfy following relations: $ \hat{\sigma}_{\mathbf{r}_{\perp}}^{2} = \hat{\sigma}_{\varphi}^{2} = \hat{\sigma}_{z}^{2} = \hat{\mathbb{I}}_{2} $ and 
$ \hat{\sigma}_{\mathbf{r}_{\perp}} \hat{\sigma}_{\varphi} = i\hat{\sigma}_{z} $, 
$ \hat{\sigma}_{\varphi} \hat{\sigma}_{z} = i\hat{\sigma}_{\mathbf{r}_{\perp}} $, 
$ \hat{\sigma}_{z} \hat{\sigma}_{\mathbf{r}_{\perp}} = i\hat{\sigma}_{\varphi} $.

Kinetic energy operator  of the electron motion transverse to the chain is represented as 
\begin{equation}
	\label{trans-kin} 
	c\bm{\hat{\sigma}} \cdot \mathbf{\hat{p}}_{\perp} = 
	c \hat{\sigma}_{\mathbf{r}_{\perp}}^{2}\bm{\hat{\sigma}} \cdot \mathbf{\hat{p}}_{\perp} = 
	\hat{\sigma}_{\mathbf{r}_{\perp}} c \left( \bm{\hat{\sigma}} \cdot \mathbf{e}_{\mathbf{r}_{\perp}} \right) \left( \bm{\hat{\sigma}} \cdot \mathbf{\hat{p}}_{\perp} \right) = 
	\hat{\sigma}_{\mathbf{r}_{\perp}} c \left( \hat{p}_{r_{\perp}} + \frac{i}{r_{\perp}} \hat{\sigma}_{z} \hat{j}_{z} \right)
\end{equation}
where 
\begin{equation}
	\label{p_rho} 
	\hat{p}_{r_{\perp}}  = \frac{1}{2} \left( \mathbf{e}_{\mathbf{r}_{\perp}} \cdot \mathbf{p}_{\perp} + \mathbf{p}_{\perp} \cdot \mathbf{e}_{\mathbf{r}_{\perp}} \right) = 
	- i\hbar \left( \frac{\partial}{\partial r_{\perp}} + \frac{1}{2r_{\perp}} \right) = - i\hbar \frac{1}{\sqrt{r_{\perp}}} \frac{\partial}{\partial r_{\perp}} \sqrt{r_{\perp}}  
\end{equation} 
is the Hermitian operator of the momentum projection on the direction $ \mathbf{r}_{\perp} $, and $ \hat{j}_{z} $ is the diagonal block of $ \hat{J}_{z} $. 

In view of the explicit expression of the operator $ \hat{p}_{r_{\perp}} $ (\ref{p_rho}), it is suitable to use substitution $ f(r_{\perp}) = (1/\sqrt{r_{\perp}}) F(r_{\perp}) $ for functions $ f_{i} $ in spinors (\ref{comps1}). This gives $ \chi^{(u/d)} = (1/\sqrt{r_{\perp}}) \Phi^{(u/d)} $, and, thus,  $ \hat{p}_{r_{\perp}} \chi^{(u/d)} = - i\hbar r_{\perp}^{-1/2} \left( \partial \Phi^{(u/d)}/\partial r_{\perp} \right) $.  The DE (\ref{DsysEq_1}) with the potential (\ref{V_eff}) with account of constants of motion $ \hat{p}_{z} $. $ \hat{j}_{z} $  and Eqs. (\ref{trans-kin}), (\ref{sigma_rho_phi_z}), takes the form
\begin{equation}
	\label{DsysEq_2} 
	\begin{array}{c} 
		- i c\hbar \hat{\sigma}_{\mathbf{r}_{\perp}} \frac{\partial \Phi^{(u)}}{\partial r_{\perp}} 
		+ \frac{c\hbar M}{r_{\perp}} \hat{\sigma}_{\varphi} \Phi^{(u)} + c\hbar k \hat{\sigma}_{z} \Phi^{(u)} + \left( - \frac{C_{eff}e^{2}Z}{r_{\perp}} - mc^{2} - E \right)  \Phi^{(d)} = 0 , \\ 
		- i c\hbar \hat{\sigma}_{\mathbf{r}_{\perp}} \frac{\partial \Phi^{(d)}}{\partial r_{\perp}} 
		+ \frac{c\hbar M}{r_{\perp}} \hat{\sigma}_{\varphi} \Phi^{(d)} + c\hbar k \hat{\sigma}_{z} \Phi^{(d)} + \left( - \frac{C_{eff}e^{2}Z}{r_{\perp}} + mc^{2} - E \right)  \Phi^{(u)} = 0 ,
	\end{array}
\end{equation} 

Adding and subtracting these two equations, we obtain equations for the spinors $ \Phi^{(\pm)} = \Phi^{(u)} \pm \Phi^{(d)} $ 
\begin{equation}
	\label{DsysEq_+,-} 
	\begin{array}{c}
		\hat{\Pi} \Phi^{(+)} - \left( \frac{C_{eff}e^{2}Z}{r_{\perp}} + E \right) \Phi^{(+)} + mc^{2} \Phi^{(-)} = 0 , \\ 
		\hat{\Pi} \Phi^{(-)} + \left( \frac{C_{eff}e^{2}Z}{r_{\perp}} + E \right) \Phi^{(-)} - mc^{2} \Phi^{(+)} = 0 
	\end{array}
\end{equation}
where the notation 
\begin{equation}
	\label{Pi} 
	\hat{\Pi} = - i c\hbar \hat{\sigma}_{\mathbf{r}_{\perp}} \frac{\partial }{\partial r_{\perp}} 
	+ \frac{c\hbar M}{r_{\perp}} \hat{\sigma}_{\varphi} + c\hbar k \hat{\sigma}_{z} 
\end{equation}
is introduced. System of equations (\ref{DsysEq_+,-}) can be reduced to the equations for each spinor 
\[
\begin{array}{c}
	\left\lbrace \left[ \hat{\Pi} + \left( \frac{C_{eff}e^{2}Z}{r_{\perp}} + E \right) \hat{\mathbb{I}}_{2} \right] 
	\left[ \hat{\Pi} - \left( \frac{C_{eff}e^{2}Z}{r_{\perp}} + E \right) \hat{\mathbb{I}}_{2} \right] + m^{2}c^{4} \right\rbrace \Phi^{(+)} = 0  , \\ 
	\left\lbrace \left[ \hat{\Pi} - \left( \frac{C_{eff}e^{2}Z}{r_{\perp}} + E \right) \hat{\mathbb{I}}_{2} \right]   
	\left[ \hat{\Pi} + \left( \frac{C_{eff}e^{2}Z}{r_{\perp}} + E \right)  \hat{\mathbb{I}}_{2}\right] + m^{2}c^{4} \right\rbrace  \Phi^{(-)} = 0  . 
\end{array}
\]   
Here we have the product of operators $ \left[ \hat{\Pi} \pm \left( E - V \right)\hat{\mathbb{I}}_{2} \right] \left[ \hat{\Pi} \mp \left( E - V \right) \hat{\mathbb{I}}_{2} \right] $, where $ V = V_{eff} $ (see Eq. \ref{V_eff}). It follows from explicit expression (\ref{Pi}) that
\[
\hat{\Pi}^{2} = - c^{2}\hbar^{2}  \frac{\partial^{2} }{\partial r_{\perp}^{2}}  
+  \frac{c^{2}\hbar^{2} M^{2}}{r_{\perp}^{2}} 
+ c^{2}\hbar^{2} k^{2}  
- \frac{c^{2}\hbar^{2} M}{r_{\perp}^{2}} \hat{\sigma}_{z} , \quad  
\left[ V , \hat{\Pi} \right] = i c\hbar \frac{\partial V }{\partial r_{\perp}} \hat{\sigma}_{\mathbf{r}_{\perp}} , 
\]
which gives us equations for the spinors $ \Phi^{(+)} $ and $ \Phi^{(-)} $:  
\begin{equation}
	\label{EqsPhi(+,-)gen} 
	\begin{array}{c} 
		\left\lbrace  - c^{2}\hbar^{2}  \frac{\partial^{2} }{\partial r_{\perp}^{2}}  
		+  \frac{c^{2}\hbar^{2} M^{2}}{r_{\perp}^{2}} 
		- \frac{c^{2}\hbar^{2} M}{r_{\perp}^{2}} \hat{\sigma}_{z} 
		- i c\hbar \frac{\partial V }{\partial r_{\perp}} \hat{\sigma}_{\mathbf{r}_{\perp}} 
		- V^{2} + 2EV - E^{2} + \mathcal{E}_{k}^{2} \right\rbrace \Phi^{(+)} = 0 , \\
		\left\lbrace  - c^{2}\hbar^{2}  \frac{\partial^{2} }{\partial r_{\perp}^{2}}  
		+  \frac{c^{2}\hbar^{2} M^{2}}{r_{\perp}^{2}} 
		- \frac{c^{2}\hbar^{2} M}{r_{\perp}^{2}} \hat{\sigma}_{z} 
		+ i c\hbar \frac{\partial V }{\partial r_{\perp}} \hat{\sigma}_{\mathbf{r}_{\perp}} 
		- V^{2} + 2EV - E^{2} + \mathcal{E}_{k}^{2} \right\rbrace \Phi^{(-)} = 0 , 
	\end{array}
\end{equation} 
where 
\begin{equation}
	\label{matcalE(k)} 
	\mathcal{E}_{k}^{2} = m^{2}c^{4} + c^{2} \hbar^{2} k^{2} .
\end{equation} 
These equations differ  by one term $ \pm i c\hbar \left( \partial V /\partial r_{\perp} \right) \hat{\sigma}_{\mathbf{r}_{\perp}} $, only. Introducing operator $ \hat{S}_{-} $ transforms spinor $ \Phi^{(+)} $ to $  \Phi^{(-)} $ and operator $ \hat{S}_{+} $ transforms spinor $ \Phi^{(-)} $ to $ \Phi^{(+)} $: 
\begin{equation}
	\label{Stransf} 
	\hat{S}_{-} \Phi^{(+)} = \Phi^{(-)} , \quad \hat{S}_{+} \Phi^{(-)} = \Phi^{(+)} , \quad 
	\hat{S}_{-} \hat{S}_{+} = \hat{S}_{+} \hat{S}_{-} = \hat{\mathbb{I}}_{2} ,
\end{equation}
and, hence, equation for $ \Phi^{(+)} $ transforms into one for $ \Phi^{(-)} $ and vice versa. This requires fulfillment of the following equalities 
\[ 
\begin{array}{c} 
	\hat{S}_{-} \left( - \frac{c^{2}\hbar^{2} M}{r_{\perp}^{2}} \hat{\sigma}_{z} 
	+ i c\hbar \frac{\partial V }{\partial r_{\perp}} \hat{\sigma}_{\mathbf{r}_{\perp}} \right) \hat{S}_{+} = - \frac{c^{2}\hbar^{2} M}{r_{\perp}^{2}} \hat{\sigma}_{z} 
	- i c\hbar \frac{\partial V }{\partial r_{\perp}} \hat{\sigma}_{\mathbf{r}_{\perp}} , \\
	\hat{S}_{+} \left( - \frac{c^{2}\hbar^{2} M}{r_{\perp}^{2}} \hat{\sigma}_{z} 
	- i c\hbar \frac{\partial V }{\partial r_{\perp}} \hat{\sigma}_{\mathbf{r}_{\perp}} \right) \hat{S}_{-} = - \frac{c^{2}\hbar^{2} M}{r_{\perp}^{2}} \hat{\sigma}_{z} 
	+ i c\hbar \frac{\partial V }{\partial r_{\perp}} \hat{\sigma}_{\mathbf{r}_{\perp}} . 
\end{array}
\] 
It follows from this requirement and from relations $ \hat{\sigma}_{z} \hat{\sigma}_{z} \hat{\sigma}_{z} = \hat{\sigma}_{z} $ and $ \hat{\sigma}_{z} \hat{\sigma}_{\mathbf{r}_{\perp}} \hat{\sigma}_{z} = - \hat{\sigma}_{\mathbf{r}_{\perp}} $ that 
\begin{equation}
	\label{S_+,-} 
	\hat{S}_{+} = A^{(+)} \hat{\sigma}_{z} , \quad \hat{S}_{-} = A^{(-)} \hat{\sigma}_{z} , \quad 
	A^{(+)} A^{(-)} = 1 .
\end{equation}
These operators allow us to represent the system of  spinor equations (\ref{DsysEq_+,-}) as two separate first order differential equations for spinors $ \Phi^{(+)} $ and $ \Phi^{(-)} $, rather than two second order differential equations (\ref{EqsPhi(+,-)gen}), by substituting $ mc^{2} \Phi^{(-)} = mc^{2} \hat{S}_{-} \Phi^{(+)} $ and $ mc^{2} \Phi^{(+)} = mc^{2} \hat{S}_{+} \Phi^{(-)} $. 

Using explicit expression (\ref{Pi}) and action of the Pauli matrices $ \hat{\sigma}_{\mathbf{r}_{\perp}} $, $ \hat{\sigma}_{\varphi} $, $ \hat{\sigma}_{z} $ on the spinors 
\[ 
\hat{\sigma}_{\bm{\mathbf{r}_{\perp}} } \Phi 
=\left( \begin{array}{c} 
	e^{-i\varphi /2} G (r_{\perp}) \\ 
	e^{i\varphi /2} F (r_{\perp}) \end{array} \right) , \quad 
\hat{\sigma}_{\varphi} \Phi = 
= \left( \begin{array}{c} 
	-i e^{-i\varphi /2} G (r_{\perp}) \\ 
	i e^{i\varphi /2} F (r_{\perp}) \end{array} \right), \quad 	
\hat{\sigma}_{z} \Phi = 
\left( \begin{array}{c} 
	e^{-i\varphi /2} F (r_{\perp}) \\ 
	- e^{i\varphi /2} G (r_{\perp}) \end{array} \right) 
\] 
(here for simplicity indices $ \pm $ are omitted), each spinor equation in Eqs. (\ref{DsysEq_+,-}) can be rewritten as a system of two equations for spinor components 
\begin{equation}
	\label{Phi^+} 
	\begin{array}{c} 
		- i c\hbar \frac{d F^{(+)} }{d r_{\perp}} 
		+ i \frac{c\hbar M}{r_{\perp}} F^{(+)} 
		- \left( \frac{C_{eff}e^{2}Z}{r_{\perp}} + E + c\hbar k + mc^{2} A^{(-)} \right) G^{(+)} = 0  ,  \\    
		- i c\hbar \frac{d G^{(+)} }{d r_{\perp}} 
		- i \frac{c\hbar M}{r_{\perp}} G^{(+)} 
		- \left( \frac{C_{eff}e^{2}Z}{r_{\perp}} + E - c\hbar k -  mc^{2} A^{(-)} \right) F^{(+)} = 0  , 
	\end{array} 
\end{equation}
and 
\begin{equation}
	\label{Phi^-} 
	\begin{array}{c} 
		- i c\hbar \frac{d F^{(-)} }{d r_{\perp}} 
		+ i \frac{c\hbar M}{r_{\perp}} F^{(-)} 
		+ \left( \frac{C_{eff}e^{2}Z}{r_{\perp}} + E - c\hbar k + mc^{2} A^{(+)} \right) G^{(-)} = 0  , \\   
		- i c\hbar \frac{d G^{(-)} }{d r_{\perp}} 
		- i \frac{c\hbar M}{r_{\perp}} G^{(-)} 
		+ \left( \frac{C_{eff}e^{2}Z}{r_{\perp}} + E + c\hbar k - mc^{2} A^{(+)} \right) F^{(-)} = 0  .  
	\end{array} 
\end{equation}

At large distances $ r_{\perp} \rightarrow \infty $ the potential and term  $ 1/r_{\perp} $ vanish: $ V \left( r_{\perp} \right) \rightarrow 0 $, $ 1/r_{\perp} \rightarrow 0 $, and,  according to Eq. (\ref{DsysEq_2}), the asymptotic behavior of spinors $ \Phi^{(u/d)} $ is determined by the following equations 
\[
\begin{array}{c}
\left( - i c\hbar \hat{\sigma}_{\mathbf{r}_{\perp}} \frac{\partial }{\partial r_{\perp}} 
+ c\hbar k \hat{\sigma}_{z} \right) \Phi^{(u)} - \left( mc^{2} + E \right) \Phi^{(d)} = 0 , \\
\left( - i c\hbar \hat{\sigma}_{\mathbf{r}_{\perp}} \frac{\partial }{\partial r_{\perp}} 
 + c\hbar k \hat{\sigma}_{z} \right) \Phi^{(d)} + \left( mc^{2} - E \right) \Phi^{(u)} = 0 , 
\end{array} 
\] 
from which it follows that each spinor satisfies the equation 
\begin{equation}
\label{asymtEqs} 
c^{2}\hbar^{2} \frac{d^{2}\Phi^{(u/d)}}{dr_{\perp}^{2}} = \left( \mathcal{E}_{k}^{2} - E^{2} \right) \Phi^{(u/d)} , \quad  \mathcal{E}_{k}^{2} = m^{2}c^{4} + c^{2} \hbar^{2} k^{2} .
\end{equation}
According to the latter equation, the asymptotic behavior of spinor components is $ f_{i}\left( r_{\perp} \right)  = \exp (- \varkappa_{\perp} r_{\perp}) f_{i} $ where
\begin{equation}
\label{varkappa} 
 \varkappa_{\perp} =  
\frac{1}{\hbar c} \sqrt{\mathcal{E}_{k}^{2} - E^{2} }  . 
\end{equation} 
 Asymptotics of the bound electron states  considered below, must be $ \Phi^{(u/d)} \rightarrow 0 $ at $ r_{\perp} \rightarrow \infty $. This takes place at $ E^{2} < \mathcal{E}_{k}^{2} $ when (\ref{varkappa}) is a real number. The energy of electrons bound by a charged string is 
\begin{equation}
\label{E_b} 
E^{2} = \mathcal{E}_{k}^{2} -\hbar^{2}  c^{2} \varkappa_{\perp}^{2} , \quad 
E = \sqrt{\mathcal{E}_{k}^{2} - \hbar^{2}c^{2} \varkappa_{\perp}^{2}} 
\end{equation} 
where $ \mathcal{E}_{k} $ is determined in Eq. (\ref{matcalE(k)}). 

So the asymptotic behavior of radial functions in spinors $ \Phi^{(u/d)} $ for bound electrons is $ \sim \exp (- \varkappa_{\perp} r_{\perp}) $ and, therefore,  the radial functions in spinors $ \Phi^{(\pm)} $ have the same behavior. Asymptotically at $ r_{\perp} \rightarrow \infty $ at  $ \sim 1/r_{\perp} \rightarrow 0 $. Equations (\ref{Phi^+}) and (\ref{Phi^-}) become 
\[
\begin{array}{c}
\hbar c \frac{d F^{(+)} }{d r_{\perp}} 
+ \left[  E + \left( mc^{2} A^{(-)} + \hbar ck \right) \right]  G^{(+)} = 0 , \\ 
\hbar c \frac{d G^{(+)} }{d r_{\perp}}   
 - \left[ E - \left[ mc^{2} A^{(-)} + \hbar c k \right) \right] F^{(+)} = 0  , 
\end{array} 
\] 
and, respectively, 
\[
\begin{array}{c} 
\hbar c\frac{d F^{(-)}}{d r_{\perp}}     
- \left[  E + \left( mc^{2} A^{(+)} - \hbar c k \right) \right] G^{(-)} = 0 , \\ 
 \hbar c\frac{d G^{(-)} }{d r_{\perp}}  
+ \left[ E - \left( mc^{2} A^{(+)} - \hbar ck \right) \right] F^{(-)}  = 0 .   
\end{array} 
\] 

After reducing the system to the second order equations for each function and substituting their asymptotic behavior $ \sim \exp (- \varkappa_{\perp} r_{\perp}) $, these equations take the form 
\[ 
\begin{array}{c}
\left\lbrace \hbar^{2}c^{2} \varkappa_{\perp}^{2} + \left[ E^{2} - \left( mc^{2} A^{(-)} + \hbar ck \right)^{2} \right] \right\rbrace F^{(+)} = 0 , \\
\\
\left\lbrace \hbar^{2}c^{2} \varkappa_{\perp}^{2} + \left[ E^{2} - \left( mc^{2} A^{(+)} - \hbar ck \right)^{2} \right] \right\rbrace F^{(-)} = 0 .
\end{array} 
\] 
from which with account of Eq. (\ref{varkappa}) the relations follow
\[
\left\lbrace \mathcal{E}_{k}^{2} - \left( mc^{2} A^{(-)} + \hbar ck \right)^{2} \right\rbrace F^{(+)} = 0 , \quad 
\left\lbrace \mathcal{E}_{k}^{2} - \left( mc^{2} A^{(+)} - \hbar ck \right)^{2} \right\rbrace F^{(-)} = 0 
\]
from which we get
\[
 mc^{2} A^{(-)} + \hbar ck = \pm \mathcal{E}_{k} , \quad mc^{2} A^{(+)} - \hbar ck = \pm \mathcal{E}_{k} .
\] 
Therefore, the constants have to satisfy the relation  $ A^{(+)} A^{(-)} = 1 $ in addition to the requirement (\ref{S_+,-}), and, depending on the sign $ \pm $, we have two possibilities for constants $ A^{(\pm)} $:   
\begin{equation}
\label{A_+,-} 
\begin{array}{ccc}
\rm I.  & 
A^{(+)}_{+} = \frac{\mathcal{E}_{k} + \hbar ck}{mc^{2}},  \quad 
A^{(-)}_{+} = \frac{\mathcal{E}_{k} - \hbar c k}{mc^{2}}  & \text{for sign $ (+) $}. \\ 
\\
\rm II.  & 
A^{(+)}_{-} = - \frac{\mathcal{E}_{k} - \hbar ck}{mc^{2}} , \quad 
\,\,\, A^{(-)}_{-} = - \frac{\mathcal{E}_{k} + \hbar ck}{mc^{2}} & \text{for sign $ (-) $}.
\end{array} 
\end{equation} 

\section{Eigen bispinors of the Dirac equation} 

\subsection{Bispinor for sign $\rm{(+)}$} 

Let us prescribe the lower index $ (+) $ to the solution and substitute expressions $ A^{(\pm)} = A^{(\pm)}_{+} $ from Eq. (\ref{A_+,-}) into Eqs. (\ref{Phi^+})-(\ref{Phi^-}). This gives us two systems of equations for the bispinor components    
\begin{equation}
\label{Phi^+_+} 
 \begin{array}{c} 
- i \hbar c\frac{d F_{+}^{(+)} }{d r_{\perp}}  
+ i \frac{\hbar cM}{r_{\perp}} F_{+}^{(+)} 
- \frac{C_{eff}e^{2}Z}{r_{\perp}} G_{+}^{(+)} - \left( \mathcal{E}_{k} + E \right) G_{+}^{(+)} = 0  ,  \\    
- i \hbar c\frac{d G_{+}^{(+)} }{d r_{\perp}} 
- i \frac{\hbar cM}{r_{\perp}} G_{+}^{(+)} 
- \frac{C_{eff}e^{2}Z}{r_{\perp}} F_{+}^{(+)} + \left( \mathcal{E}_{k} - E \right) F_{+}^{(+)} = 0  , 
\end{array} 
\end{equation} 
\begin{equation}
\label{Phi^-_+} 
\begin{array}{c} 
- i \hbar c\frac{d F_{+}^{(-)} }{d r_{\perp}} 
+ i \frac{\hbar cM}{r_{\perp}} F_{+}^{(-)} 
+ \frac{C_{eff}e^{2}Z}{r_{\perp}} G_{+}^{(-)} + \left( \mathcal{E}_{k} + E \right) G_{+}^{(-)} = 0  , \\   
- i \hbar c \frac{d G_{+}^{(-)} }{d r_{\perp}} 
- i \frac{\hbar cM}{r_{\perp}} G_{+}^{(-)} 
+ \frac{C_{eff}e^{2}Z}{r_{\perp}} F_{+}^{(-)} - \left( \mathcal{E}_{k} - E \right) F_{+}^{(-)} = 0  .  
\end{array}  
\end{equation}

To find their solution, the following substitution can be useful:
\begin{equation}
\label{subsF,G_+} 
F_{+}^{(\pm)} =\frac{1}{2}  e^{- \varkappa_{\perp} r_{\perp}} r_{\perp}^{\gamma} d^{-1/2} 
\left( u_{+}^{\pm} + v_{+}^{\pm} \right) , \quad G_{+}^{(\pm)} = 
\frac{1}{2}i e^{- \varkappa_{\perp} r_{\perp}} r_{\perp}^{\gamma} d^{1/2}  \left( u_{+}^{\pm} - v_{+}^{\pm} \right) ,   
\end{equation} 
 where 
\begin{equation}
\label{d} 
d = \sqrt{\frac{\mathcal{E}_{k} - E}{\mathcal{E}_{k} + E}} = \frac{\hbar c\varkappa_{\perp}}{\mathcal{E}_{k} + E} = \frac{\mathcal{E}_{k} - E}{\hbar c\varkappa_{\perp} } ,
\end{equation} 
This gives us systems of equations for functions $ u^{(\pm)}_{+} $ and $ v^{(\pm)}_{+} $: 
\begin{equation}
\label{Eqs-u,v_+^+} 
\begin{array}{c}
 r_{\perp} \frac{d u_{+}^{(+)} }{d r_{\perp}}   
- \left( \frac{ C_{eff}\alpha Z}{\hbar c\varkappa_{\perp}} E - \gamma \right) u_{+}^{(+)} 
- \left( M + \frac{ C_{eff}\alpha Z}{\hbar c\varkappa_{\perp}} \mathcal{E}_{k} \right) v_{+}^{(+)} = 0 ,  \\    
 r_{\perp} \frac{d v_{+}^{(+)} }{d r_{\perp}}    
 + \left( \frac{ C_{eff}\alpha Z}{\hbar c\varkappa_{\perp}} E + \gamma 
 - 2 \varkappa_{\perp} r_{\perp} \right) v_{+}^{(+)}    
- \left( M - \frac{ C_{eff}\alpha Z}{\hbar c\varkappa_{\perp}} \mathcal{E}_{k} \right) u_{+}^{(+)} = 0  ,  
\end{array}
\end{equation} 
and 
\begin{equation}
\label{Eqs-u,v_+^-} 
\begin{array}{c} 
 r_{\perp} \frac{d u_{+}^{(-)} }{d r_{\perp}}    
 + \left( \frac{ C_{eff}\alpha Z}{\hbar c\varkappa_{\perp}} E + \gamma 
 - 2 \varkappa_{\perp} r_{\perp} \right) u_{+}^{(-)}    
- \left( M - \frac{ C_{eff}\alpha Z}{\hbar c\varkappa_{\perp}} \mathcal{E}_{k} \right) v_{+}^{(-)} = 0  , \\  
 r_{\perp} \frac{d v_{+}^{(-)} }{d r_{\perp}}   
- \left( \frac{ C_{eff}\alpha Z}{\hbar c\varkappa_{\perp}} E - \gamma \right) v_{+}^{(-)} 
- \left( M + \frac{ C_{eff}\alpha Z}{\hbar c\varkappa_{\perp}} \mathcal{E}_{k} \right) u_{+}^{(-)} = 0 , 
\end{array} 
\end{equation}
where $ \alpha = e^{2} /\hbar c $ is the Sommerfeld fine structure constant. 

Note that Eqs. (\ref{Eqs-u,v_+^-}) match Eqs. (\ref{Eqs-u,v_+^+}) with the change of functions $ u_{+}^{(+)} \rightarrow v_{+}^{(-)} $ and $ v_{+}^{(+)} \rightarrow u_{+}^{(-)} $, and, accordingly, it is enough to search the solution of one system, only, for example, of Eqs. (\ref{Eqs-u,v_+^+}). After obtaining solutions $ u_{+}^{(+)},\,v_{+}^{(+)} $, we can write down the solution of Eqs. (\ref{Eqs-u,v_+^-}) as $ u_{+}^{(-)} = Av_{+}^{(+)} $ and $ v_{+}^{(-)} = Au_{+}^{(+)} $. 

As it has been shown above, two first order Eqs. (\ref{Eqs-u,v_+^+}) can be reduced to the second order equations for each function:  
\begin{equation}
\label{Eq_u,Eq_v} 
\begin{array}{c}
  r_{\perp}^{2} \frac{d^{2} u^{(+)} }{d r_{\perp}^{2}}  
+ \left( 1 + 2 \gamma - 2 \varkappa_{\perp} r_{\perp} \right) r_{\perp} \frac{d u^{(+)} }{d r_{\perp}} 
 + 2 \varkappa_{\perp} r_{\perp} \left( \frac{ C_{eff}\alpha Z}{\hbar c \varkappa_{\perp}} E - \gamma \right) u^{(+)} + \\
 + \left( \gamma^{2} - M^{2} + C_{eff}^{2}\alpha^{2} Z^{2} \right) u^{(+)} = 0 , \\
 \\
  r_{\perp}^{2} \frac{d^{2} v^{(+)} }{d r_{\perp}^{2}}   
 + \left( 1 + 2 \gamma - 2 \varkappa_{\perp} r_{\perp} \right) r_{\perp} \frac{d v^{(+)} }{d r_{\perp}} 
 + 2 \varkappa_{\perp} r_{\perp} \left( \frac{ C_{eff}\alpha Z}{\hbar c \varkappa_{\perp}} E -\gamma - 1 \right) v^{(+)}  + \\
 + \left( \gamma^{2}  - M^{2} + C_{eff}^{2}\alpha^{2} Z^{2} \right) v^{(+)} = 0 .    
\end{array} 
\end{equation}
 
From here it follows that functions $ u^{(+)} $ and $ v^{(+)} $ satisfy the hypergeometric differential equations. The solution has to be finite throughout the space and, therefore, the  hypergeometric series should terminate for some value of $ n_{r} $. This condition leads to the equalities 
\begin{equation}
\label{cond-gamma,E} 
\gamma^2 = M^{2} - C_{eff}^{2}Z^{2}\alpha^{2}  \quad \text{and} \quad 
\frac{C_{eff}\alpha ZE}{\hbar c\varkappa_{\perp}} - \gamma = n_{r} 
\end{equation}
where $ n_{r} $ is a positive integer number. Under this condition  two independent Eqs. (\ref{Eq_u,Eq_v}) become 
\begin{equation}
\label{LaguerEq} 
\begin{array}{c} 
  \xi \frac{d^{2} u^{(+)}_{+} }{d \xi^{2}}  
+ \left( 1 + 2 \gamma - \xi \right) \frac{d u^{(+)}_{+} }{d \xi} + n_{r} u^{(+)}_{+} = 0 , \\  
  \xi \frac{d^{2} v^{(+)}_{+} }{d \xi^{2}}  
+ \left( 1 + 2 \gamma - \xi \right) \frac{d v^{(+)}_{+} }{d \xi} + (n_{r} - 1) v^{(+)}_{+} = 0  
\end{array}
\end{equation} 
where the dimensionless variable is introduced
\begin{equation}
	\label{xi}
	\xi = 2 \varkappa _{\perp} r_{\perp}.
\end{equation} 
From here it follows that functions $ u^{(+)}_{+}(\xi) $ and $ v^{(+)}_{+}(\xi) $ satisfy the equation known as the equation for the Laguerre polynomials $ \mathit{L}_{n}^{2\gamma} \left( \xi \right) $, and, therefore, the solutions are 
\begin{equation}
 \label{u,v^+,-} 
\begin{array}{c}
 u_{+}^{(+)} = A_{1} \mathit{L}_{n_{r}}^{2\gamma} \left( \xi \right)  \quad \text{and}  \quad 
 v_{+}^{(+)} = B_{1} \mathit{L}_{n_{r}-1}^{2\gamma} \left( \xi \right) , \\
 u_{+}^{(-)} = B_{2} \mathit{L}_{n-1}^{2\gamma} \left( \xi \right)  \quad \text{and}  \quad  
v_{+}^{(-)} = A_{2} \mathit{L}_{n}^{2\gamma} \left( \xi \right) .   
\end{array} 
 \end{equation} 
 
Functions $ u_{+}^{(+)} $ and $ v_{+}^{(+)} $, as well as $ u_{+}^{(-)} $ and $ v_{+}^{(-)} $, are not independent because at values of $ E $ determined from conditions (\ref{cond-gamma,E}), they ought to be the solutions of Eqs. (\ref{Eqs-u,v_+^+}) and (\ref{Eqs-u,v_+^-}), respectively. 

The first equality in Eq. (\ref{cond-gamma,E}) gives  us
\begin{equation}
\label{gamma} 
\gamma = \sqrt{M^{2} - C_{eff}^{2} Z^{2} \alpha^{2} }
\end{equation}
where the square root has the positive sign, only, as it follows from the condition of the normalization integral convergence. From the second equality in Eq. (\ref{cond-gamma,E})  we have 
\begin{equation}
\label{kappa_n,M} 
\hbar c\varkappa_{\perp} = \frac{C_{eff} Z \alpha \mathcal{E}_{k}}{\sqrt{\left( \gamma + n_{r} \right)^{2} + C_{eff}^{2} Z^{2}\alpha^{2} }} = \frac{C_{eff} Z \alpha \mathcal{E}_{k}}{\sqrt{ M^{2} + n_{r}^{2} + 2\gamma n_{r} }}
\end{equation}
and, therefore, the energy eigenvalue (\ref{E_b}) is given by the expression
\begin{equation}
\label{E_n,M,k} 
E_{n,M,k} = \sqrt{\mathcal{E}_{k}^{2} - \frac{C_{eff}^{2} Z^{2}\alpha^{2} \mathcal{E}_{k}^{2}}{\left( \gamma + n_{r} \right)^{2} + C_{eff}^{2} Z^{2}\alpha^{2}}} = 
\sqrt{\mathcal{E}_{k}^{2} - \frac{C_{eff}^{2} Z^{2}\alpha^{2} \mathcal{E}_{k}^{2}}{ M^{2} + n_{r}^{2} + 2\gamma n_{r} } }.
\end{equation} 

After transition to variable (\ref{xi}) with account of functional relations of Laguerre polynomials, it follows  from Eqs. (\ref{Eqs-u,v_+^+}) that at the $\gamma $-values  (\ref{gamma})-(\ref{E_n,M,k}) the coefficients $ A $ and $ B $ are connected by the relation 
\begin{equation}
\label{A_1,B_1} 
\frac{B_{1}}{A_{1}} = \frac{B_{2}}{A_{2}} = - \frac{n_{r} + 2\gamma}{ \sqrt{M^{2} + n_{r} \left( n_{r} + 2\gamma \right) } + M } = - \frac{q_{-}}{q_{+}} \sqrt{\frac{2\gamma + n_{r}}{n_{r}}}   
\end{equation}
where 
\begin{equation}
\label{q_pm} 
q_{\pm} \equiv q_{\pm}\left( M,n_{r} \right) = \left( \sqrt{M^{2} + n_{r} \left( n_{r} + 2\gamma \right) } \pm M \right)^{1/2} .
\end{equation} 
 
Expressions (\ref{subsF,G_+}) and (\ref{u,v^+,-}) determine radial functions in the eigen bispinors of the DE (\ref{DsysEq_1}). To write these functions, we will use normalized orthogonal Laguerre polynomials 
\begin{equation}
\label{ortnormLpol} 
\mathcal{L}_{n}^{2\gamma} \left( \xi \right) = \sqrt{\frac{n!}{\Gamma \left( 2\gamma + n + 1 \right)}}  \mathit{L}_{n}^{2\gamma} \left( \xi \right) , \quad 
\int_{0}^{\infty} e^{-\xi} \xi^{2\gamma} \mathcal{L}_{n}^{2\gamma} \left( \xi \right) \mathcal{L}_{m}^{2\gamma} \left( \xi \right) d\xi = \delta_{n,m} .
\end{equation}
As a result, the eigen bispinor which corresponds to the sign $ (+) $ in Eq. (\ref{A_+,-}), is 
\begin{equation}
\label{Psi_+,1} 
\Psi_{k,n_{r},M,+} \left( \mathbf{r} \right) = \frac{1}{2} e^{i \left( kz + M\varphi \right) } e^{- \varkappa_{\perp} r_{\perp}} r_{\perp}^{\gamma - \frac{1}{2}} 
\left( \begin{array}{c} \tilde{\Phi}_{n_{r},+}^{(u)}  \\ \tilde{\Phi}_{n_{r},+}^{(d)} 
\end{array} \right) 
\end{equation}
with 
\begin{equation}
\label{Phi*ud,1} 
\tilde{\Phi}_{n_{r},+}^{(u)} = \left( \begin{array}{c} 
C_{1,+} e^{-i\varphi /2} d^{-1/2} P_{n_{r},+}\left( \xi \right) \\ 
i C_{2,+} e^{i\varphi /2} d^{1/2} Q_{n_{r},+}\left( \xi \right) \end{array} \right) ,  \quad  
\tilde{\Phi}_{n_{r},+}^{(d)} = \left( \begin{array}{c} 
C_{2,+} e^{-i\varphi /2} d^{-1/2} P_{n_{r},+}\left( \xi \right) \\ 
i C_{1,+} e^{i\varphi /2} d^{1/2} Q_{n_{r},+}\left( \xi \right)  \end{array} \right) . 
\end{equation}
where polynomial notations are introduced: 
\begin{equation}
\label{polynoms_+} 
P_{n_{r},+}\left( \xi \right) = q_{+} \mathcal{L}_{n_{r}}^{2\gamma} \left( \xi \right) 
        - q_{-} \mathcal{L}_{n_{r}-1}^{2\gamma} \left( \xi \right) , \quad 
Q_{n_{r},+}\left( \xi \right) = q_{+} \mathcal{L}_{n_{r}}^{2\gamma} \left( \xi \right) 
+ q_{-} \mathcal{L}_{n_{r}-1}^{2\gamma} \left( \xi \right) .
\end{equation}
Here $ C_{1,+} , \: C_{2,+} $ are arbitrary constants which satisfy the normalization condition. 

The eigen bispinor (\ref{Psi_+,1}) describes bound electron states with the sign $ (+) $ according to the sign choice in Eq. (\ref{A_+,-}) with the set of quantum numbers $ \lbrace k,n_{r},M \rbrace $. The radial quantum number $ n_{r} = 0,1,2,\ldots $ determines the order of the polynomials (\ref{polynoms_+}) in the radial functions. At $ n_{r} = 1,2,\ldots $ ($ n_{r} \neq 0 $) the positive and negative values of total angular momentum $ M = \pm 1/2,\, \pm 3/2, \, \ldots $ around the chain are allowed. But at $ n_{r} = 0 $ Eqs. (\ref{Eqs-u,v_+^+})-(\ref{Eqs-u,v_+^-}) admit solutions $ v_{+}^{(+)} = u_{+}^{(-)} = 0 $, given in Eq. (\ref{u,v^+,-}), at positive values of $ M $, only. Namely, at $ M > 0 $ the coefficient $ q_{-}\left( M,n_{r} \right) $ in Eq. (\ref{polynoms_+}) is equal to  zero at $ n_{r} = 0 $ ($ q_{-}\left( M,0 \right) = 0 $) and $ \mathcal{L}_{-1}^{2\gamma} $ with the negative index does not appear in Eq. (\ref{polynoms_+}). 
Therefore, at $ n_{r} = 0 $ the eigen bispinor (\ref{Psi_+,1}) is $ \Psi_{k,0,M>0,+} \left( \mathbf{r} \right) $ in which $ M $ takes positive values, only.

\subsection{Bispinor for sign $\rm{(-)}$} 

At sign $ (-) $ for constants $ A^{(\pm)} $ we substitute expressions $ A^{(\pm)} = A^{(\pm)}_{-} $ from Eq. (\ref{A_+,-}) into Eqs. (\ref{Phi^+})-(\ref{Phi^-}) and obtain the system of equations for the spinor components in $ \Phi^{(\pm)}_{-} $  
\begin{equation}
\label{Phi^+_-} 
\begin{array}{c} 
- i \hbar c\frac{d F_{-}^{(+)} }{d r_{\perp}} 
+ i \frac{\hbar c M}{r_{\perp}} F_{-}^{(+)} 
- \frac{C_{eff}e^{2}Z}{r_{\perp}} G_{-}^{(+)} + \left( \mathcal{E}_{k} - E \right) G_{-}^{(+)} = 0  ,  \\     
- i \hbar c\frac{d G_{-}^{(+)} }{d r_{\perp}} - i \frac{\hbar cM}{r_{\perp}} G_{-}^{(+)} 
- \frac{C_{eff}e^{2}Z}{r_{\perp}} F_{-}^{(+)} - \left( \mathcal{E}_{k} + E \right) F_{-}^{(+)} = 0 , 
\end{array} 
\end{equation}
and 
\begin{equation}
\label{Phi^-_-} 
\begin{array}{c} 
- i \hbar c\frac{d F_{-}^{(-)} }{d r_{\perp}} 
+ i \frac{\hbar c M}{r_{\perp}} F_{-}^{(-)} 
+ \frac{C_{eff}e^{2}Z}{r_{\perp}} G_{-}^{(-)} - \left( \mathcal{E}_{k} - E \right) G_{-}^{(-)} = 0  , \\   
- i \hbar c \frac{d G_{-}^{(-)} }{d r_{\perp}} 
- i \frac{\hbar cM}{r_{\perp}} G_{-}^{(-)} 
+ \frac{C_{eff}e^{2}Z}{r_{\perp}} F_{-}^{(-)} + \left( \mathcal{E}_{k} + E \right) F_{-}^{(-)} = 0  .  
\end{array} 
\end{equation} 

To find their solution, we use the substitution 
\begin{equation}
\label{subsF,G_-} 
F_{-}^{(\pm)} =\frac{1}{2}  i e^{- \varkappa_{\perp} r_{\perp}} r_{\perp}^{\gamma} d^{1/2}  \left( u_{-}^{(\pm)} + v_{-}^{(\pm)} \right) , \quad 
G_{-}^{(\pm)} = \frac{1}{2} e^{- \varkappa_{\perp} r_{\perp}} r_{\perp}^{\gamma} d^{-1/2} \left( u_{-}^{(\pm)} - v_{-}^{(\pm)} \right)  ,   
\end{equation} 
which transforms Eqs. (\ref{Phi^+_-})-(\ref{Phi^-_-}) into equations for functions $ u^{(\pm)}_{-} $ and $ v^{(\pm)}_{-} $. For 	$ u^{(+)}_{-} $ and $ v^{(+)}_{-} $ we have
\begin{equation}
\label{syst-u,v^(+)_-} 
\begin{array}{c}
r_{\perp} \frac{du_{-}^{(+)} }{dr_{\perp}} 
- \left( \frac{E C_{eff}Z\alpha}{\hbar c\varkappa_{\perp}} - \gamma \right) u_{-}^{(+)} 
- \left( M - \frac{\mathcal{E}_{k}C_{eff}Z\alpha}{\hbar c\varkappa_{\perp}} \right) v_{-}^{(+)} = 0 ,  \\  
r_{\perp} \frac{dv_{-}^{(+)} }{dr_{\perp}} 
+ \left( \frac{EC_{eff}Z\alpha}{\hbar c\varkappa_{\perp}} + \gamma - 2\varkappa_{\perp} r_{\perp} \right) v_{-}^{(+)} - \left( M + \frac{C_{eff}Z\alpha \mathcal{E}_{k}}{\hbar c\varkappa_{\perp}} \right) u_{-}^{(+)} = 0 .
\end{array}
\end{equation} 
As above, for functions $ u^{(-)}_{-} $ and $ v^{(-)}_{-} $ similar  equations are obtained: 
\[
\begin{array}{c} 
r_{\perp} \frac{du_{-}^{(-)} }{dr_{\perp}} 
+ \left( \frac{C_{eff}Z\alpha E}{\hbar c\varkappa_{\perp}} + \gamma - 2\varkappa_{\perp} r_{\perp} \right) u_{-}^{(-)} - \left( M + \frac{C_{eff}Z\alpha \mathcal{E}_{k}}{\hbar c\varkappa_{\perp}} \right) v_{-}^{(-)} = 0 ,  \\  
r_{\perp} \frac{dv_{-}^{(-)} }{dr_{\perp}} 
- \left( \frac{ C_{eff}Z\alpha E}{\hbar c\varkappa_{\perp}} - \gamma \right) v_{-}^{(-)} 
- \left( M - \frac{C_{eff}Z\alpha \mathcal{E}_{k}}{\hbar c\varkappa_{\perp}} \right) u_{-}^{(-)} = 0 ,  
\end{array}
\] 
with the functions change $ u_{-}^{(+)} \rightarrow v_{-}^{(-)} $ and $ v_{-}^{(+)} \rightarrow u_{-}^{(-)} $. 

Similar to Eqs. (\ref{Eqs-u,v_+^+}) and (\ref{Eqs-u,v_+^-}), solutions of the latter  equations at conditions (\ref{cond-gamma,E}) are expressed via Laguerre polynomials 
\begin{equation}
\label{u,v^+.-_-} 
u_{-}^{(+)} = A_{3} \mathit{L}_{n_{r}}^{2\gamma} , \quad 
v_{-}^{(+)} = B_{3} \mathit{L}_{n_{r}-1}^{2\gamma} ; \quad \text{and} \quad
 u_{-}^{(-)} = B_{4} \mathit{L}_{n_{r}-1}^{2\gamma} ,  \quad  
v_{-}^{(-)} = A_{4} \mathit{L}_{n_{r}}^{2\gamma}  
\end{equation} 
where constants $ A_{3(4)} $ and $ B_{3(4)} $ are connected by relation 
\begin{equation}
\label{B3/A3} 
\frac{B_{3}}{A_{3}} = \frac{B_{4}}{A_{4}} = \frac{q_{+}}{q_{-}} \sqrt{\frac{n + 2\gamma}{n}} 
\end{equation}
and $ q_{\pm} $ is determined in Eq. (\ref{q_pm}). 

Thus, the eigen bispinor with quantum numbers $ (k,n_{r},M,- )$ is 
\begin{equation}
\label{Psi_-,1} 
\Psi_{k,n_{r},M,-} \left( \mathbf{r} \right) = \frac{1}{2} e^{i \left( kz + M\varphi \right) } e^{- \varkappa_{\perp} r_{\perp}} r_{\perp}^{\gamma - \frac{1}{2}} 
\left( \begin{array}{c} \tilde{\Phi}_{n_{r},-}^{(u)}  \\ \tilde{\Phi}_{n_{r},-}^{(d)} 
\end{array} \right) 
\end{equation} 
with 
\begin{equation}
\label{Phi^ud_-} 
\tilde{\Phi}_{n_{r},-}^{(u)} = \left( \begin{array}{c}  
i e^{-i\varphi /2} d^{1/2} C_{1,-} P_{n_{r},-}\left( \xi \right)  \\  
e^{i\varphi /2} d^{-1/2} C_{2,-} Q_{n_{r},-}\left( \xi \right) \end{array} \right) , \quad 
\tilde{\Phi}_{n_{r},-}^{(d)} = \left( \begin{array}{c} 
i e^{-i\varphi /2} d^{1/2} C_{2,-} P_{n_{r},-}\left( \xi \right)  \\ 
e^{i\varphi /2} d^{-1/2} C_{1,-} Q_{n_{r},-}\left( \xi \right)
\end{array} \right)
\end{equation}
where 
\begin{equation}
\label{polynoms_-} 
P_{n_{r},-}\left( \xi \right) = q_{-} \mathcal{L}_{n_{r}}^{2\gamma} \left( \xi \right)
+ q_{+} \mathcal{L}_{n_{r}-1}^{2\gamma} \left( \xi \right) , \quad 
Q_{n_{r},-}\left( \xi \right) = q_{-} \mathcal{L}_{n_{r}}^{2\gamma} \left( \xi \right)
- q_{+} \mathcal{L}_{n_{r}-1}^{2\gamma} \left( \xi \right) . 
\end{equation} 

For the radial quantum number $ n_{r} = 1,2,\ldots $ positive and negative values of $ M $ are allowed and the coefficient in front of $ \mathcal{L}_{-1}^{2\gamma} $ in Eq. (\ref{polynoms_+}) vanishes, $ q_{+}\left( M,0 \right) = 0 $,  at $ n_{r} = 0 $ for $ M < 0 $, only. This indicates that the eigen bispinor (\ref{Psi_-,1}) which corresponds to the choice of the sign $ (-) $ in Eq. (\ref{A_+,-}), at $ n_{r} = 0 $ becomes $ \Psi_{k,0,M<0,-} $ with negative values of $ M $, only.

\subsection{The system of the orthonormalized eigen bispinors} 

Solutions of the DE (\ref{DsysEq_1}) give the system of the eigen bispinors $ \Psi_{k,n_{r},M,\sigma} $ (see Eqs.  (\ref{Psi_+,1}) and (\ref{Psi_-,1}) ) which describe stationary states of electrons bound by the charged atomic chain. Stationary states are characterized by the set of quantum numbers $ \lbrace k,n_{r},M,\sigma \rbrace $. Here the wave vector $ k $ determines the kinetic momentum $ \hbar k $ along the chain, positive and negative half-integer values $ M = \pm 1/2,\, \pm 3/2, \, \ldots $ determine the total angular momentum $ \hbar M $ around the chain ($ \hat{J}_{z} $), the radial quantum number $ n_{r} = 0,1,2,\ldots $ indicating the order of polynomials (\ref{polynoms_+}) in the radial functions, accounts for the allowed discrete energy levels, and the two-valued number $ \sigma = \pm 1 $ can be considered as the fourth quantum number, namely as the spin quantum number.

Complex constants $ C_{1,\pm} $ and $ C_{2,\pm} $ in the bispinors satisfy the normalization and orthogonality conditions 
\[ 
\int \Psi_{k,n_{r},M,\sigma}^{\dagger} \left( \mathbf{r} \right) \Psi_{k,n_{r},M,\sigma'} \left( \mathbf{r} \right) dV = \delta_{\sigma ,\sigma'} . 
\] 
After integration and taking into account expressions (\ref{d}), (\ref{q_pm}), and (\ref{gamma})-(\ref{E_n,M,k}), the normalization condition gives us
\[  
 \vert C_{1} \vert^{2} + \vert C_{2} \vert^{2} = C^{2} = \frac{2(2\varkappa_{\perp} )^{2\gamma + 1}}{L \pi \left( d + d^{-1} \right) \left( q_{+}^{2} + q_{-}^{2} \right)} =  
 \frac{(2\varkappa_{\perp} )^{2\gamma + 1} C_{eff} Z \alpha }{2\pi L \left( M^{2} + n_{r}^{2} + 2\gamma n_{r} \right)}   .
\]
Here constants $ C_{1/2,\pm} $ can be represented in the general form $ C_{1,\pm} = a_{\pm} C $ and $ C_{2,\pm} = b_{\pm} C $ with $ \vert a_{\pm} \vert^{2} + \vert b_{\pm} \vert^{2} = 1 $. 

The orthogonality condition for bispinors $ \Psi_{k,n,M,+} $ and $ \Psi_{k,n,M,-} $ requires that $ b_{-} = a_{+} \equiv a $ and $ a_{-} = b_{+} \equiv b $. Then, considering the normalization condition with the requirement $ \vert a \vert^{2} + \vert b \vert^{2} = 1 $ and equalities 
\[ 
\begin{array}{c}
\frac{\sqrt{C_{eff} Z_{v}\alpha}}{ \sqrt{M^{2} + n_{r}^{2} + 2\gamma n_{r}} } d^{1/2} = 
\frac{\sqrt{2}}{\left(  M^{2} + n_{r}^{2} + 2\gamma n_{r} \right)^{1/4} } 
\sqrt{\frac{\mathcal{E}_{k} - E}{ 2\mathcal{E}_{k} } } , \\
\frac{\sqrt{C_{eff} Z_{v}\alpha}}{ \sqrt{M^{2} + n_{r}^{2} + 2\gamma n_{r}} } d^{-1/2} = 
\frac{\sqrt{2}}{ \left( M^{2} + n_{r}^{2} + 2\gamma n_{r} \right)^{1/4}} 
\sqrt{\frac{\mathcal{E}_{k} + E}{ 2\mathcal{E}_{k} } } ,
\end{array}  \quad  E = E_{k,n_{r},M} , 
\] 
we introduce notations  
\begin{equation}
\label{notations} 
a = e^{i\phi_{1}} \cos \frac{\theta}{2} , \; b = e^{i\phi_{2}} \sin \frac{\theta}{2} , \quad 
\sqrt{\frac{\mathcal{E}_{k} + E_{k,n_{r},M}}{2\mathcal{E}_{k}}} = \cos \frac{\tau}{2} , \; 
\sqrt{\frac{\mathcal{E}_{k} - E_{k,n_{r},M}}{2\mathcal{E}_{k}}} = \sin \frac{\tau}{2}  
\end{equation} 
for the system of orthonormalized bispinors describing quasi-1D electrons in the atomic chain. 

For $ n_{r} = 0 $ we have two bispinors $ \Psi_{k,0,M>0,+} $ and $ \Psi_{k,0,M<0,-} $ which can be written as 
\begin{equation}
\label{bispin_n=0} 
\Psi_{k,0,\sigma \vert M \vert,\sigma} = e^{i\phi_{0}/2} 
\sqrt{\frac{(2\varkappa_{\perp} )^{2\gamma + 1} }{2\pi L \Gamma \left( 2\gamma + 1 \right) } } 
e^{i \left( kz + \sigma \vert M \vert \varphi \right) } e^{- \varkappa_{\perp} r_{\perp}} r_{\perp}^{\gamma - \frac{1}{2}} 
\left( \begin{array}{c} \chi_{\sigma}^{(u)}  \\ \chi_{\sigma}^{(d)} 
\end{array} \right) ,
\end{equation}
where 
\begin{equation}
\label{chi^ud_+} 
\chi_{+}^{(u)} = \left( \begin{array}{c} 
\cos \frac{\theta}{2} \cos \frac{\tau}{2} e^{-i\left( \varphi + \phi \right) /2} \\  
i \sin \frac{\theta}{2} \sin \frac{\tau}{2} e^{i\left( \varphi + \phi \right) /2}   
\end{array} \right) ,  \quad  
\chi_{+}^{(d)} = \left( \begin{array}{c} 
\sin \frac{\theta}{2} \cos \frac{\tau}{2} e^{-i\left( \varphi - \phi \right) /2} \\  
i \cos \frac{\theta}{2} \sin \frac{\tau}{2} e^{i\left( \varphi - \phi \right) /2}  
 \end{array} \right) , 
\end{equation}
and 
\begin{equation}
\label{chi^U,d_-} 
\chi_{-}^{(u)} = \left( \begin{array}{c}  
i \sin \frac{\theta}{2} \sin \frac{\tau}{2} e^{-i(\varphi + \phi )/2} \\    
\cos \frac{\theta}{2} \cos \frac{\tau}{2} e^{i(\varphi + \phi )/2} 
\end{array} \right) , \quad  
\chi_{-}^{(d)} = \left( \begin{array}{c} 
i \cos \frac{\theta}{2} \sin \frac{\tau}{2} e^{-i(\varphi - \phi )/2}  \\    
\sin \frac{\theta}{2} \cos \frac{\tau}{2} e^{i(\varphi - \phi ) /2} 
\end{array} \right) . 
\end{equation}

For $ n_{r} = 1,2,\ldots $ the bispinors are 
\begin{equation}
\label{Psi_k,n,M,s} 
\Psi_{k,n_{r},M,\sigma} = e^{i\phi_{0}/2} \sqrt{\frac{(2\varkappa_{\perp} )^{2\gamma + 1} }{4\pi L \sqrt{M^{2} + n_{r}^{2} + 2\gamma n_{r}} } } e^{i \left( kz + M \varphi \right) } e^{- \varkappa_{\perp} r_{\perp}} r_{\perp}^{\gamma - \frac{1}{2}} 
\left( \begin{array}{c} \psi_{n_{r},\sigma}^{(u)}  \\ \psi_{n_{r},\sigma}^{(d)} 
\end{array} \right) ,  
\end{equation}
where the upper and lower spinors are given by  relations
\begin{equation}
 \label{psi^u,d_+} 
\psi_{n_{r},+}^{(u)} =  \left( \begin{array}{c} 
\cos \frac{\theta}{2} \cos \frac{\tau}{2} e^{-i\left( \varphi + \phi \right) /2} 
P_{n_{r},+} \\ 
i \sin \frac{\theta}{2} \sin \frac{\tau}{2} e^{i\left( \varphi + \phi \right) /2} 
Q_{n_{r},+} 
\end{array} \right) , \; 
\psi_{n_{r},+}^{(d)} = \left( \begin{array}{c} 
\sin \frac{\theta}{2} \cos \frac{\tau}{2} e^{-i\left( \varphi - \phi \right) /2} 
P_{n_{r},+}  \\ 
i \cos \frac{\theta}{2} \sin \frac{\tau}{2} e^{i\left( \varphi - \phi \right) /2} 
Q_{n_{r},+} 
 \end{array} \right)  
 \end{equation} 
and 
\begin{equation}
\label{psi^u,d_-} 
\psi_{n,-}^{(u)} = \left( \begin{array}{c}  
i \sin \frac{\theta}{2} \sin \frac{\tau}{2} e^{-i(\varphi + \phi )/2} 
P_{n_{r},-}  \\     
\cos \frac{\theta}{2} \cos \frac{\tau}{2} e^{i(\varphi + \phi )/2} 
Q_{n_{r},-} \end{array} \right) , \; 
\psi_{n,-}^{(d)} = \left( \begin{array}{c} 
i \cos \frac{\theta}{2} \sin \frac{\tau}{2} e^{-i(\varphi - \phi )/2} 
P_{n_{r},-}  \\    
\sin \frac{\theta}{2} \cos \frac{\tau}{2} e^{i(\varphi - \phi ) /2} 
Q_{n_{r},-} \end{array} \right) . 
\end{equation}
Here polynomials $ P_{n_{r},\pm} = P_{n_{r},\pm}\left( \xi \right) $ and $ Q_{n_{r},\pm} = Q_{n_{r},\pm}\left( \xi \right) $ are determined  in Eqs. (\ref{polynoms_+}) and (\ref{polynoms_-}). The  $ \tau $ parameter is small ($ \sin (\tau /2) \sim C_{eff} Z \alpha \ll 1 $), and two parameters $ 0 \leq \theta \leq \pi $ and $ 0 \leq \phi \leq 2\pi $ are the variables of the spin degree of freedom which determine spin orientation $ \langle \mathbf{s} \rangle_{k,n_{r},M,\sigma} $ in the given state $ \Psi_{k,n_{r},M,\sigma} $.  We remind that $ \mathbf{s} = (\hbar/2) \bm{\hat{\Sigma}} $.

\section{Conclusion} 

Thus, we have shown that stationary states of electrons bound by the charged chain, are described by the eigen bispinors $ \Psi_{k,n_{r},M,\sigma} $ of the DE and are characterized by the full set of quantum numbers $ \lbrace k,n_{r},M,\sigma \rbrace $. The corresponding eigenvalues (\ref{E_n,M,k}) are given by expressions
\begin{equation}
	\label{disp}
E_{k,n_{r},M} = \sqrt{\mathcal{E}_{k}^{2} - \frac{C_{eff}^{2} Z^{2}\alpha^{2} \mathcal{E}_{k}^{2}}{ M^{2} + n_{r}^{2} + 2\gamma n_{r} } } .
\end{equation}
Here $ k $ is the wave vector which describes 1D electron motion along the chain with the energy band dispersion $ E_{k,n_{r},M} $  corresponding to the discrete values of numbers $ \vert M \vert $ and $ n_{r} $. These bands can be enumerated by introducing the principal quantum number analogous to the hydrogen-like spectrum. 
For this let us represent the half-integer number $ \vert M \vert = 1/2,\, 3/2 ,\, \ldots $ in the form $ \vert M \vert = \ell + 1/2 $ where the number $ \ell $ takes values $ \ell = 0, 1,2,\ldots $. This allows us to write down the equality
\[
 M^{2} + n_{r}^{2} + 2\gamma n_{r} = \left( n_{r} + \ell + \frac{1}{2} \right)^{2} - 2 \left( \vert M \vert - \gamma \right) n_{r} ,
\] 
and to introduce the \textit{principal quantum number} 
\begin{equation}
\label{princ_q_n} 
n = 1 + n_{r} + \ell 
\end{equation}  
which takes values $ n = 1,2,\ldots $. The following relation is valid:
\[
 M^{2} + n_{r}^{2} + 2\gamma n_{r} = \left( n - \frac{1}{2} \right)^{2} - \Delta \left( n_{r},\ell \right) 
\]
where notation is used
\begin{equation}
\label{Delta_n} 
 \Delta \left( n_{r},\ell  \right) = 2 \left( \vert M \vert - \gamma \right) n_{r} =  
 2 C_{eff}^{2} Z^{2} \alpha^{2} \frac{ n_{r}}{\ell + 1/2 + \gamma}  . 
\end{equation} 
As it will be shown below, this value plays the role of the band splitting parameter. 
The energy spectrum (\ref{E_n,M,k}) reads now as
\begin{equation}
\label{E_n,k} 
E_{k,n,M} = \sqrt{\mathcal{E}_{k}^{2} - \frac{C_{eff}^{2} Z^{2}\alpha^{2} \mathcal{E}_{k}^{2}}{ \left( n - 1/2 \right)^{2} - \Delta \left( n_{r},\ell \right) } } .
\end{equation}


At non-relativistic values of energy $ c^{2} \hbar^{2} k^{2} \ll m^{2}c^{4} $ the energy dispersion can be approximated as 
\[
E_{n,M,k} = \sqrt{m^{2}c^{4} + c^{2} \hbar^{2} k^{2} 
 - \frac{C_{eff}^{2} Z^{2}\alpha^{2} m^{2}c^{4}}{ \left( n - 1/2 \right)^{2} - \Delta \left( n_{r},\ell \right) } } \simeq 
 mc^{2} + \frac{ \hbar^{2} k^{2}}{2m} + \mathcal{E}_{n,\ell} + \ldots 
\] 
where 
\begin{equation}
\label{nonrelE} 
\mathcal{E}_{n,\ell} = - \frac{1}{2} \frac{C_{eff}^{2} Z_{v}^{2}\alpha^{2} mc^{2} }{ \left( n - 1/2 \right)^{2} - \Delta \left( n_{r},\ell \right) } \simeq 
 - \frac{1}{2} \frac{C_{eff}^{2} Z_{v}^{2}\mathcal{E}_h }{ \left( n + 1/2 \right)^{2}} \left( 1 + \frac{\Delta \left( n_{r},\ell \right)}{\left( n - 1/2 \right)^{2}} \right)  
\end{equation}
is the energy spectrum of quasi-1D electrons bound by monatomic chain. Here $\mathcal{E}_h$ is the hartree  fundamental atomic unit f energy 
\[
\mathcal{E}_h = \alpha^{2} mc^{2} = \frac{m e^{4}}{\hbar^{2}} = 27.21 \text{eV} . 
\]

Thus, we have shown that stationary states of bound electrons in atomic chain can be characterized by the principal quantum number $ n = 1,2,\ldots $ which describes the so called  "gross structure" of the energy bands, according to Eq.  (\ref{E_n,k} ). At the given $ n $ the radial quantum and orbital numbers take values $ n_{r} = 0,1,\ldots ,n-1 $ and  $ \ell = n - 1 - n_{r} $, respectively. Due to the presence of the term $ \Delta \left( n_{r},\ell \right) $  in Eqs. (\ref{E_n,k})-(\ref{nonrelE}), the spectral band with the principal quantum number $ n $ splits into $ n $ subbands forming the fine structure characterized by the the band splitting parameter (\ref{Delta_n}). 
The scale of the fine structure splitting relative to the gross structure energies is of the order of $ \Delta \left( n_{r},\ell \right) \sim (Z\alpha)^{2} $ and is given  below for some values of $n$. 
\begin{equation}
	\label{fine-str}
	\begin{array}{ccc} 
		n & \text{States} \: \vert k,n_{r},M,\sigma \rangle & \text{Energy} \: E_{n,\vert M \vert}(k) = \mathcal{E}_{n,\vert M \vert} + \frac{ \hbar^{2} k^{2}}{2m}	
		\\  
		1 & \text{two states} \: \vert k,0,\sigma 1/2,\sigma \rangle  & \mathcal{E}_{1,1/2} = - 2\mathcal{E}_h C_{eff}^{2} Z^{2} \\ 
		\\
		2 & \begin{array}{c}
			\text{four states} \: \vert k,1,\pm 1/2,\pm \rangle \\
			\text{two states} \: \vert k,0,\sigma 3/2,\sigma \rangle
		\end{array} 
		& \begin{array}{c} 
			\mathcal{E}_{2,1/2} = - \frac{2}{9} \mathcal{E}_h C_{eff}^{2} Z^{2} \left( 1 + \frac{8}{9}  C_{eff}^{2} Z^{2} \alpha^{2} \right)  \\ 		
			\mathcal{E}_{2,3/2} = - \frac{2}{9} \mathcal{E}_h C_{eff}^{2} Z^{2} 
		\end{array} \\
		\\ 		
		3 & \begin{array}{c}
			\text{four states} \: \vert k,2,\pm 1/2,\pm \rangle \\ 		
			\text{four states} \: \vert k,1,\pm 3/2,\pm \rangle \\ 		
			\text{two states} \: \vert k,0,\sigma 5/2,\sigma \rangle
		\end{array} 
		& \begin{array}{c} 
			\mathcal{E}_{3,1/2} = - \frac{2}{25} \mathcal{E}_h C_{eff}^{2} Z^{2} \left( 1 + \frac{16}{25}  C_{eff}^{2} Z^{2} \alpha^{2} \right)  \\ 		
			\mathcal{E}_{3,3/2} = - \frac{2}{25} \mathcal{E}_h C_{eff}^{2} Z^{2} \left( 1 + \frac{8}{75}  C_{eff}^{2} Z^{2} \alpha^{2} \right)  \\ 				
			\mathcal{E}_{3,5/2} = - \frac{2}{25} \mathcal{E}_h C_{eff}^{2} Z^{2}	 		
		\end{array} 
	\end{array}
\end{equation}

\section{Acknowledgements}
This work has been done under the support of the grant 2025.07/0335 of the Fundamental Research Foundation of Ukraine.

\end{document}